\documentclass[a4paper]{jpconf}
\usepackage{graphicx}
\newcommand{\bm}[1]{ \mbox{\boldmath $#1$}  }

\begin{document}

\title{Decay of low-lying $^{12}$C resonances within a 3$\alpha$ cluster 
model}
\author{R \'Alvarez-Rodr\'{\i}guez$^1$, A S Jensen$^1$, D V Fedorov$^1$, 
H O U Fynbo$^1$ and E Garrido$^2$}

\address{ $^1$ Department of Physics and Astronomy, University of Aarhus 
DK-8000 Aarhus C, Denmark\\  
$^2$ Instituto de Estructura de la Materia, Consejo Superior de 
Investigaciones Cient\'{\i}ficas E-28006 Madrid, Spain}

\ead{raquel@phys.au.dk}

\begin{abstract}
We compute energy distributions of three $\alpha$-particles emerging from
the decay of $^{12}$C resonances by means of the hyperspherical adiabatic 
expansion method combined with complex scaling. The
large distance continuum properties of the wave functions are crucial
and must be accurately calculated. The substantial changes from small
to large distances determine the decay mechanisms. We illustrate by
computing the energy distributions from decays of the $1^{+}$ and 
$3^-$-resonances in $^{12}$C. These states are dominated by
direct and sequential decays into the three-body continuum respectively.
\end{abstract}

\section{Introduction.}

The low-lying resonance states of $^{12}$C have been studied over many years
both theoretically and experimentally, motivated, in part, for its
astrophysical importance \cite{fynbo05}. 
Surprisingly, the energies and structure of the
low-lying resonances below 16~MeV are still not well known.

We investigate in this contribution the decay
of low-lying resonance states into three particle final states for the
case of $^{12}$C, assuming that the decay mechanism is independent 
of how the initial state was formed. We describe the decay in analogy
with $\alpha$-decay, assuming that the three fragments are formed before 
entering the barrier at sufficiently small distances to allow the 
three-body treatment. The hyperradius $\rho$ provides 
a measure of distances for our three-body problem.
Outside the range of the strong interaction,
only the Coulomb and centrifugal barriers remain, since we have assumed 
that the small-distance many-body dynamics is unimportant for the process.

The same kind of three-body techniques could be
used in other astrophysical interesting processes, e.g., the 
triple alpha process ($3\alpha \to^{12}$C$ + \gamma$), the formation
of $^9$Be ($2\alpha +  n \to^{9}$Be$ + \gamma$), or the 2 proton capture
involved in the rp-process \cite{gri05}.

By measuring the properties of the particles after the decay, we get
double information, as for example the energy distributions and
information 
about the decay mechanism, e.g. direct or sequential \cite{alv07b}.
The theoretical information about distributions of relative energies is 
contained in the large distance part of the resonance wave function.
Numerically converged results in the appropriate region of 
$\rho$-values are then needed in order to have a reliable computation.

\section{Theoretical framework.}

The decay of a $^{12}$C resonance into 3 $\alpha$ particles is a pure
three-body problem of nuclear physics. Therefore we describe
$^{12}$C as a 3$\alpha$-cluster system. Moreover triple-$\alpha$
decay is the only open non-electromagnetic decay channel for this nucleus.
We use the Faddeev equations and
solve them in coordinate space using the adiabatic hyperspherical expansion 
method \cite{nie01}. In order to obtain the resonances we use the complex 
scaling method. The 
so-called hyperradius $\rho$ is the most important of the coordinates, and 
is defined as
\begin{eqnarray}  
  m_N  \rho^{2} =  \frac{m_{\alpha}}{3} \sum^{3}_{i<j}
 \left(\bm{r}_{i}-\bm{r}_j\right)^{2} = 
  m_{\alpha}  \sum^{3}_{i=1} 
 \left(\bm{r}_{i}-\bm{R}\right)^{2}  \label{e120} \;,
\end{eqnarray}
for three identical particles of mass $m_\alpha = 4m_N$, where 
we choose $m_N$ to be equal to the nucleon mass, 
$\bm{r}_{i}$ is the coordinate of the i-th particle and $\bm{R}$ 
is the centre-of-mass coordinate. 
 
We first must determine the interaction $V_i$ reproducing the low-energy
two-body scattering properties. In this case, we have chosen an Ali-Bodmer
potential \cite{ali66} slightly modified in order to reproduce 
the s-wave resonance of $^8$Be. We add then a three-body 
potential, whose range corresponds 
to three touching $\alpha$-particles. These potentials are chosen 
independently for each $J^\pi$.

\begin{figure}
\vspace*{-2.0cm}
\begin{center}
\includegraphics[width=25pc,angle=-90]{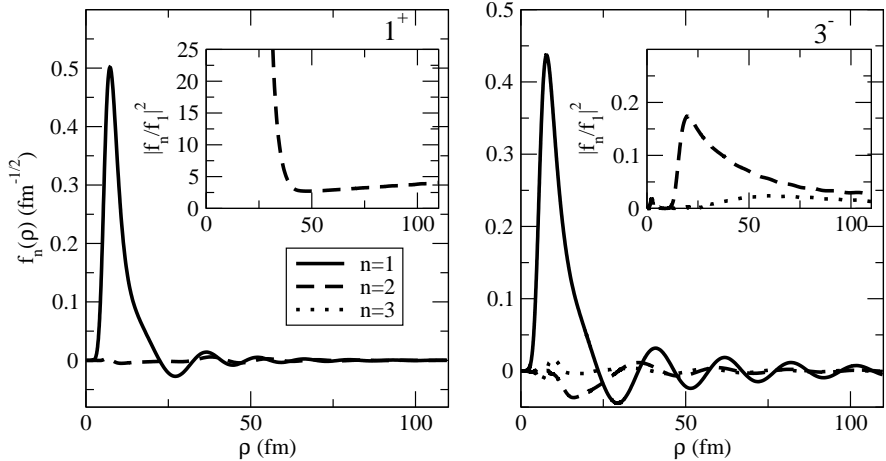}
\end{center}
\vspace*{-1.6cm}
\caption{\label{figr} The adiabatic radial wave functions and the ratios
of the different components (inset) for the $1^+$ (left) and $3^-$
(right) resonances of $^{12}$C at an excitation energy of
12.7~MeV and 9.6~MeV respectively.   }
\end{figure}

The total wave function is expanded on the angular 
eigenfunctions $\Phi_{nJM} (\rho,\Omega)$ obtained as solutions to the Faddeev 
equations for fixed $\rho$,
\begin{equation}
\Psi^{JM} = \frac{1}{\rho^{5/2}}\sum_n f_n (\rho) \Phi_{nJM} (\rho,\Omega)\;,
\end{equation}
where the radial coefficients $f_n (\rho)$ are obtained from
the coupled set of radial equations. The effective adiabatic potentials
are the eigenvalues of the angular part.

At intermediate
distances the potential has a barrier that determines the resonance width;
and at large distances the resonance wave functions contain information about
distributions of relative energies.

The asymptotic behaviour of the decaying resonance wave function determines
the energy distribution in the observable final state. This energy 
distribution can be computed in coordinate space, except for a 
phase-space factor, as the integral of the absolute square of the 
total wave function for a large value of the hyperradius \cite{gar07a}. 
We shall 
explore the conjecture that the final state can be obtained entirely
within the present cluster model.

\section{Results.}

\begin{figure}
\begin{center}
\includegraphics[width=35pc]{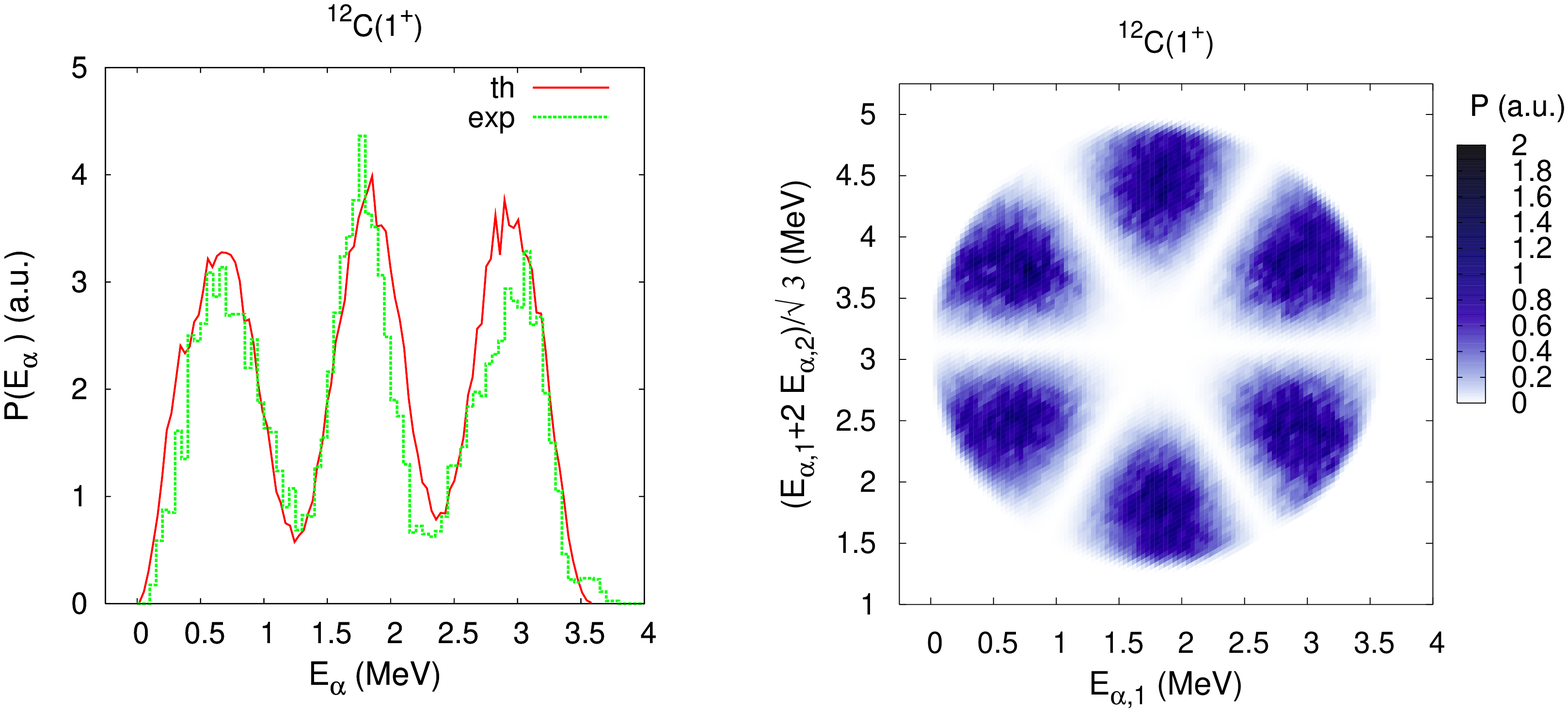}
\end{center}
\caption{\label{fig1p} The $\alpha$-particle energy distribution (left)
and Dalitz plot (right) for the $1^+$ resonance of $^{12}$C at 5.43~MeV 
above the 3$\alpha$
threshold at an excitation energy of 12.7~MeV. Experimental data are
from \cite{dig06}.}
\end{figure}

By using this method we could investigate a number of the low-lying 
$^{12}$C resonances below 15~MeV, i.e., two $0^+$, three $2^+$, two $4^+$, 
and one of each of $1^\pm$, $2^-$, $3^\pm$,
$4^-$ and $6^+$ \cite{alv07}. 

In fig. \ref{figr} we have plotted the radial wave functions together
with the absolute squared values of the ratios between the different 
components for both $1^+$ and $3^-$ resonances of $^{12}$C. 
One can observe that these ratios are
fairly insensitive to variations of the hyperradius at large distances, that
is where the energy distributions are computed. This means that for
those distances the asymptotical behaviour has been already reached.

Fig.\ref{fig1p} shows the individual $\alpha$-particle
energy distribution and the Dalitz plot for the $1^+$ state of 
$^{12}$C at 12.7~MeV of excitation energy (or at 5.43~MeV above 
the 3$\alpha$ threshold). This $1^+$ state of $^{12}$C is 
often referred to as a shell-model
state, which means that it can not be described as a cluster structure.
Should the final state consists of 3$\alpha$ particles,
a three-body description is unavoidable. Moreover, due to 
parity considerations the sequential decay via the $^8$Be ground state is
forbidden. We have compared the theoretical energy distribution
with the accurate experimental data for this resonance populated in
$\beta$-decay \cite{dig06}. 
Fig. \ref{fig1p} shows an impressive agreement between theory
and experiment. If we compare the Dalitz plot with fig. 
1 of the reference \cite{fyn03}, it is clear that we are able to reproduce
also the basic features of the Dalitz plot obtained from the experimental data.

\begin{figure}
\vspace*{-0.7cm}
\begin{center}
\includegraphics[width=20pc,angle=-90]{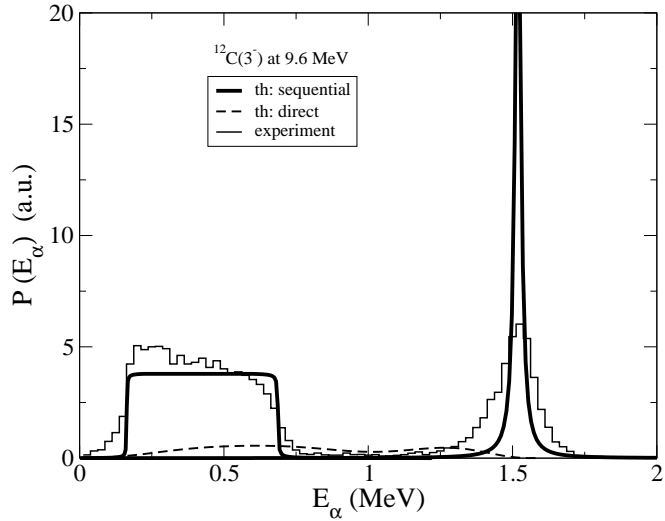}
\end{center}
\vspace*{-0.6cm}
\caption{\label{fig3m} The $\alpha$-particle energy distribution for the
$3^-$ resonance of $^{12}$C at 2.33~MeV 
above the 3$\alpha$ threshold at excitation energy of 9.6~MeV. Contributions
from direct and sequential decays via $^8$Be(g.s.) are shown. Experimental
data are from \cite{oli}.}
\end{figure}

The energy distribution for the natural parity state $3^-$ 
at 9.6~MeV of excitation energy (or at 2.33~MeV above the 3$\alpha$
threshold)
shows the characteristic features of a sequential decay via the $^8$Be 
ground state, i.e. a narrow high-energy peak, and a distribution 
around $E_{max}/4$ (see fig. \ref{fig3m}). Since the complex rotated
two-body asymptotic behaviour correspond to a bound state, the
computed distributions are not accurate. Instead we can exploit the fact
that precisely one of the adiabatic potentials asymptotically must describe
the two-body resonance and the third particle far away. The corresponding
sequential decay is then two consecutive two-body decays with the related 
well-known distinct kinematics. This component can then be
substituted by the Fourier transform of the known asymptotic 
two-body behaviour, as it has been previously done in \cite{alv07b}.
By looking at the ratios between the different
adiabatic radial wave functions (fig. \ref{figr}), 
we can estimate that the direct
decay is about 4\% and sequential decay via $^8$Be($0^+$) is 96\% of the
total distribution.
This sequential part of the energy distribution is approximated by 
the leading order Breit-Wigner shape for the
first emitted $\alpha$-particle. It has a high-energy peak at the most
probable position (2/3 of the resonance energy). The width is the sum of the
three-body decaying resonance width and the width of the intermediate 
two-body resonance. By kinematic conditions we can compute the energy of
the two $\alpha$-particles emerging from $^8$Be, 
that gives rise to the peak at lower energy.
After removing the contribution from the first adiabatic potential, the 
remaining energy distribution is described as direct decay by the computed
three-body continuum coordinate space wave function.
The distribution is rather uniform but mostly visible between the 
separate peaks of the sequential decay. We have compared our result
with some preliminary data from the reaction 
$^{10}$B($^3$He,p$\alpha\alpha\alpha$) studied at CMAM (Madrid)
by O. Kirsebom and collaborators \cite{oli}. Even though the analysis
of the data is not yet finished, we find a nice agreement between the
basic features of the measured and the theoretical energy distributions.

\section{Summary and Conclusions.}

We have applied a general method to compute the particle-energy 
distributions of some of the three-body decaying $^{12}$C many-body 
resonances. We conjectured
that the energy distributions of the decay fragments are insensitive
to the initial many-body structure. The energy distributions are
then determined by the energy and three-body resonance structure
as obtained in a three-$\alpha$ cluster model. These momentum 
distributions are determined by the coordinate space
wave functions at large distances. We can separate components with
two- and three-body asymptotics that correspond to sequential and
direct decays respectively.

We have shown the examples of $1^+$ and $3^-$ states of $^{12}$C.
The $1^+$ resonance is best described by direct decay 
into the three-body continuum, whereas the $3^-$ resonance have 
substantial cluster components and decays preferentially via the $^{8}$Be 
ground state. In both cases we have compared our computation with 
the measured distributions and found a nice agreement between
theory and experiment. 

In conclusion, we predict energy distributions of particles emitted 
in three-body decays. 
Both sequential and direct, and both short and long range interactions 
are treated.

\ack
R.A.R. acknowledges support by a post-doctoral fellowship from 
Ministerio de Educaci\'on y Ciencia (Spain).

\end{document}